\documentclass[conference]{IEEEtran}
\makeatletter
\def\ps@headings{%
\def\@oddhead{\mbox{}\scriptsize\rightmark \hfil \thepage}%
\def\@evenhead{\scriptsize\thepage \hfil \leftmark\mbox{}}%
\def\@oddfoot{}%
\def\@evenfoot{}}
\makeatother
\ifCLASSINFOpdf
  \usepackage[pdftex]{graphicx}
\else
  \usepackage[dvips]{graphicx}
\fi

\usepackage[cmex10]{amsmath}
\usepackage{amssymb, amsthm}
\usepackage{graphicx}
\usepackage{url, cite, verbatim}
\usepackage[ruled,vlined,linesnumbered]{algorithm2e}

\DeclareMathOperator*{\argmin}{arg\,min}
\DeclareMathOperator*{\argmax}{arg\,max}

\usepackage{amsfonts}
\newtheorem{theorem}{Theorem}
\newtheorem{lemma}{Lemma}
\newtheorem{propo}{Proposition}

\newcommand{\ls}[1]  
   {\dimen0=\fontdimen6\the=#1\dimen0
    \advance\lineskip.5\fontdimen5\the\lineskip-\dimen0
    \lineskiplimit=.9\lineskip
    \baselineskip=\lineskip
    \advance\baselineskip\dimen0
    \normallineskip\lineskip
    \normallineskiplimit\lineskiplimit
    \normalbaselineskip\baselineskip
    \ignorespaces
   }

\usepackage[tight,footnotesize]{subfigure}

\begin{document}

\title{On Cell Association and Scheduling Policies in Femtocell Networks}

\author{\IEEEauthorblockN{Hui Zhou, Donglin Hu, Saketh Anuma Reddy, Shiwen Mao, and Prathima Agrawal}
\IEEEauthorblockA{Department of Electrical and Computer Engineering, Auburn University, Auburn, AL, USA\\
}
}


\maketitle

\ls{0.881}

\begin{abstract}
Femtocells are recognized effective for improving network coverage and capacity, and reducing power consumption due to the reduced range of wireless transmissions. Although highly appealing, a plethora of challenging problems need to be addressed for fully harvesting its potential. In this paper, we investigate the problem of cell association and service scheduling in femtocell networks. In addition to the general goal of offloading macro base station (MBS) traffic, we also aim to minimize the latency of service requested by users, while considering both open and closed access strategies. We show the cell association problem is NP-hard, and propose several near-optimal solution algorithms for assigning users to base stations (BS), including a sequential fixing algorithm, a rounding approximation algorithm, a greedy approximation algorithm, and a randomized algorithm. For the service scheduling problem, we develop an optimal algorithm to minimize the average waiting time for the users associated with the same BS. The proposed algorithms are analyzed with respect to performance bounds, approximation ratios, and optimality, and are evaluated with simulations. 
\end{abstract}


\IEEEpeerreviewmaketitle

\pagestyle{plain}\thispagestyle{plain}

\section{Introduction}

\begin{figure} [!t]
	\vspace{0.1in}
	\centering
		\includegraphics[width=2.5in]{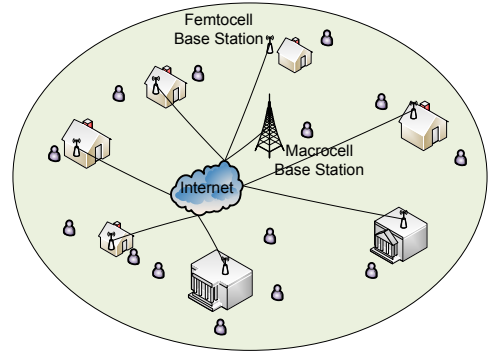}
	\caption{Illustration of a two-tier femtocell network.}
	\label{fig:femtocell}
	\vspace{-0.15in}
\end{figure}

A femtocell, as shown in Fig.~\ref{fig:femtocell}, is a relatively small cellular network with a femtocell base station (FBS), usually deployed in places where signal reception from the macro base station (MBS) is weak due to long distance or obstacles. An FBS is typically the size of a residential gateway or smaller and connects to the service provider's network via broadband connections.
FBS is designed to serve approved users within its coverage to offload wireless traffic from MBS. Due to shortened wireless transmission distance, femtocell is shown very effective in reducing transmit power and boosting signal-to-interference-plus-noise ratio (SINR), which lead to prolonged battery life of mobile devices, improved network coverage, and enhanced network capacity~\cite{Andrews12}. 

Femtocells have gained a lot of attention from both academia and industry in the recent past.  
The three largest cellular network operators in the United States (i.e., AT\&T, Sprint and Verizon) have offered commercial femtocell products and service recently. 
Although highly promising, 
a plethora of problems with both technical and economic natures have not been fully addressed yet. In~\cite{Andrews12}, a comprehensive discussion is provided of the challenging technical issues in femtocell networks, ranging from synchronization, cell association, network organization, to quality of service (QoS) provisioning.  

Unlike the MBS, whose placement is planned and optimized by operators, FBS's are usually randomly deployed by users. When the chaotic femtocell placement meets randomly distributed mobile users, cell association (or load balancing) becomes a critical problem for the performance of femtocell networks. For example, an FBS might be deployed at a place with high user density. With an inappropriate cell association strategy, this FBS may have to serve all the users within its coverage, leading to very high load at this FBS and high service latency for its users. An effective cell association scheme should be used in this case 
to evenly distribute the load among neighboring FBS's and/or MBS. 
The cell association 
problem is particularly prominent in femtocell networks due to the unreliability of FBS's. The operation of an FBS may be interrupted by its owner (e.g., turned off after office hours); 
it may also experience power outage or any other faults. 
Then all the users initially associated with this FBS should be quickly assigned to other neighboring FBS's or the MBS. 
It is a load balancing problem on how to effectively associate these users with neighboring BS's without introducing a load burst and performance degradation at a particular BS. 

In this paper, we investigate the problem of cell association and service scheduling in a two-tier femtocell network. In addition to the general goal of offloading wireless traffic from the MBS, we also aim to minimize the latency of service requested by users, while considering both open and closed access strategies. 
In particular, we consider one MBS and multiple FBS's serving randomly distributed mobile users. 
Users request to the BS's for downlink transmission of data packets. Without loss of generality, we assume that each user is allowed to connect to either the MBS or an FBS. The cell associate problem is to assign the users to the BS's such that the transmission of all the data packets can be completed as soon as possible. When multiple users are associated with one BS, we 
also aim to develop a service scheduling scheme such that the average waiting time for the users will be minimized. 

We provide a general framework for the cell association problem for both open and closed access scenarios, which can be reduced to the classic load balancing problem and is NP-hard~\cite{Kleinberg05}. Therefore, we develop effective near-optimal algorithms with guaranteed performance. In particular, we first provide a sequential fixing algorithm based on a linear programming (LP) relaxation,
which can achieve the best performance among the proposed schemes but with a relatively high computational complexity. To reduce the complexity, we propose a rounding approximation algorithm that ensures an $(\rho+1)$-approximation of the optimal solution, and a greedy approximation algorithm that ensures a $(2\rho)$-approximation of the optimal solution. 
To further reduce the requirement on frequently updated channel state information (CSI), we then develop a randomized algorithm that allows a user to randomly pick a BS to connect to from a reduced BS list. Once the reduced BS list is generated by the randomized algorithm, no information exchange is required among users. An upper bound for the maximum expected service time achieved by the randomized algorithm is then derived. 

After the users are assigned to the BS's, we next address the service scheduling problem for determining the transmission order of the data packets requested by the users associated with the same BS. We develop a simple algorithm to minimize the average waiting time for the users, and prove its optimality. 
In addition rigorous analysis of the proposed algorithms with respect to performance bounds, approximation ratios, and optimality, we also evaluate the proposed schemes with simulations, where superior performance is observed.   

The remainder of this paper is organized as follows. The related work is discussed in Section~\ref{sec:RelWork}. We present the system model in Section~\ref{sec:sysMod}. Cell Association problem formulation and solutions are presented in Section~\ref{sec:ProbSol}. The scheduling problem 
is studied in~\ref{sec:SevSchd}. The proposed algorithm are evaluated in Section~\ref{sec:PerfEva}. 
Section~\ref{sec:Conc} concludes this paper.

\section{Related Work}\label{sec:RelWork}
 
Femtocells have been acknowledged as an effective solution to the capacity problem of wireless networks. Ref.~\cite{Andrews12} provided comprehensive discussions of the technical issues, regulatory concerns, and economic incentives in femtocell networks. There are three different access control strategies in femtocell networks, open access, closed access and hybrid access. The pros and cons of these strategies were studied in~\cite{Roche10}.
 
Deploying femtocells also means introducing interference if no appropriate mitigation strategy is incorporated. 
Considerable research have been conducted on interference mitigation by assigning users to proper orthogonal channels~\cite{Hu12JSAC}.
 
Apart from the studies on interference mitigation, there are an increasing number of papers on cell association or cell selection under various scenarios~\cite{Dhahri12, Madan10, Corroy12, Zhou13, Jo12, Mukherjee12}. Dhahri and Ohtsuki in~\cite{Dhahri12} proposed a learning-based cell selection method for an open access femtocell network. The authors in~\cite{Madan10} described new paradigms
of cell association in heterogeneous networks with the help of third-party backhaul connections. Their simple and lightweight methodologies and algorithms incur very low signaling overhead. In~\cite{Corroy12}, a convex optimization problem was formulated for cell association and a dynamic range extension algorithm was proposed to maximize the minimum rate of users on the downlink of heterogeneous networks. However, this paper did not directly optimize the load balancing in Heterogeneous Networks (HetNet), but rather focused on the sum rate and min rate. In~\cite{Zhou13}, a cell association and access control scheme was presented to maximize network capacity while achieving fairness among users.
In~\cite{Jo12}, the authors provided an analytical framework for evaluating outage probability and spectral efficiency with flexible cell association in heterogeneous cellular networks.
Mukherjee in~\cite{Mukherjee12} analyzed the downlink SINR distribution in heterogeneous networks with biased cell association.
 
There are also some interesting prior work on load balancing in cellular networks. A theoretical framework was presented in~\cite{Kim12} for distributed user association and cell load balancing under spatially heterogeneous traffic distribution. A distributed $\alpha$-optimal algorithm was proposed and it supports different load-balancing objectives, which include rate-optimal, throughput-optimal, delay-optimal, and load-equalizing, as $\alpha$ is set to different values.
In~\cite{Son09}, the authors developed an off-line optimal algorithm for load balancing to achieve network-wide proportional fairness in multi-cell networks. They considered partial frequency reuse (PFR) jointly with load-balancing in a multi-cell network to achieve network-wide proportional fairness. An on-line practical algorithm was also proposed and the expected throughput was taken as the decision making metric. On-line assignments when users arrive one at a time was studied extensively in computer science literature. The competitive ratio analysis in~\cite{Azar92} showed that any deterministic on-line algorithm can achieve a competitive ratio of $\log n$, where $n$ is the number of servers.
 
We find most of the 
related research was focused on offloading MBS traffic and improving network capacity with FBS's. In the following sections, we propose several cell association and transmission scheduling schemes with the objective of minimizing service latency in femtocell networks.

\section{System Model}\label{sec:sysMod}

We consider a two-tier femtocell network with $M$ base stations: one MBS (indexed by $1$) and $M-1$ FBS's (indexed from $2$ to $M$). The All the BS's are connected to the Internet via broadband wired connections. There are $N$ mobile users randomly located within the coverage of the femtocell network. We assume the MBS and FBS's are well synchronized and they share the same spectrum. Assume each user requests a fixed-length data packet from one of the $M$ BS's. The problem is to assign the users to the BS's and schedule the transmission of their requested data packets at each BS, such that the transmissions can be finished as earlier as possible. 

\subsection{Link Capacity}

Let $P_m$ be the transmit power of BS $m$ and $G_{m,n}$ the channel gain between the BS and user $n$. According to the Shannon Theorem, the network capacity of user $n$ connected to BS $m$ is given by
\begin{eqnarray}\label{eq:Cmn}
	C_{m,n}=B\log_2\left(1+\frac{G_{m,n}P_m}{\sigma^2+I_{m,n}}\right), 
\end{eqnarray}
where $B$ is network bandwidth,\footnote{It is well-known from queuing theory that a single server single buffer queue has the lowest delay than splitting the service capacity to multiple servers or maintaining multiple queues.} 
$\sigma^2$ is noise power density, and $I_{m,n}$ is the interference from all other BS's. We have that
\begin{eqnarray}\label{eq:Imn} 
	I_{m,n}=\sum_{i=1}^M G_{i,n}P_i-G_{m,n}P_m=I_n-G_{m,n}P_m, 
\end{eqnarray}
where $I_n$ is the sum of interference from all BS's to user $n$. It does not depend on which BS user $n$ is connected to and is a constant for each user. Substituting (\ref{eq:Imn}) into (\ref{eq:Cmn}), we have
\begin{eqnarray}\label{eq:Cmn1}
	C_{m,n} &=& B\log_2\left(1+\frac{G_{m,n}P_m}{\sigma^2+I_n-G_{m,n}P_m}\right) \nonumber \\
	        &=& B\log_2\left(\frac{1}{1-\eta_{m,n}}\right), 
\end{eqnarray}
where $\eta_{m,n}$ is signal to interference plus noise ratio (SINR), the same ratio of the received power in $I_n$ at user $n$. 

\subsection{Service Time}

We assume each user requests a fixed-length data packet from one of the BS's. For simplicity of notation, we assume all the packets have the same length, denoted as $L$. Then the processing/service time of BS $m$ for user $n$ is given by
\begin{eqnarray}\label{eq:tmn}
	t_{m,n}=L/ C_{m,n}.
\end{eqnarray}
The service time depends on the link capacity $C_{m,n}$ as given in~(\ref{eq:Cmn1}). Note that the service time defined here is actually the transmission delay, i.e., the time it takes to finish the transmission of the data packet. The propagation delay is negligible due to the short distance and is ignored.

\subsection{Femtocell Access Control}

The type of access control for femtocells can be classified into two categories: closed access and open access. The open-access strategy allows all mobile users of an operator to connect to the FBS's; in this case, femtocells are often deployed 
by an operator to 
enhance coverage in an area where there is a coverage hole. With the closed access strategy, only a specific user group can get service from the FBS's~\cite{Golaup09}. Although closed access has been shown to decrease system throughput by 15\%, surveys suggest that closed access is users' favorite option~\cite{Hasan09}.  

In this paper, we consider both access strategies. Let $\mathcal{A}_m$ denote the set of users that can connect to BS $m$ and $\mathcal{B}_n$ the set of BS's that user $n$ can connect to. Both open and closed access strategies can be easily modeled by these two sets. Specifically, for open access, we have $\mathcal{A}_m=\{1,\cdots,N\}$ and $\mathcal{B}_n=\{1,\cdots,M\}$.

\section{Cell Association Problem Formulation and Proposed Schemes}\label{sec:ProbSol}

To make the complex problem tractable, we divide the problem into two steps. 
First, we assign each user to one of the $M$ BS's with the objective of minimizing the total service time on each BS. 
Second, we schedule the service order at each BS to minimize the average waiting time of users.  

\subsection{Problem Statement}

The cell association problem can be formulated as a load balancing problem. Given a set of $N$ users and a set of $M$ BS's. Each user $n$ has a service time $t_{m,n}$ if it is connected to BS $m$. Let $\mathcal{C}_m$ denote the set of users assigned to BS $m$. Then it takes a total amount of time $T_m=\sum_{n\in \mathcal{C}_m}t_{m,n}$ 
for BS $m$ to transmit all the packets. For optimal network-wide performance, we seek to minimize the maximum load among all the BS's, i.e., 
\begin{equation} \label{eq:T}
  \min \; T=\max_m \{T_m\} = \max_m \left\{\sum_{n\in \mathcal{C}_m}t_{m,n} \right\}. 
\end{equation}

We find the cell association problem is similar to a load balancing problem. However, our problem is more challenging than the classic load balancing problem, where the service time of a user is identical when connecting to any BS. 
In our cell associate problem, the service time is a function of the link capacity as in~(\ref{eq:tmn}). 
Its solution depends on not only user $n$, but also BS $m$. This cell association problem is easily seen to be NP-hard: when all the $t_{m,n}$'s are identical for any BS $m$, the problem is reduced to the classic load balancing problem, which is NP-hard~\cite{Kleinberg05}. 

In the remainder of this section, we develop effective algorithms to solve the cell association problem. In particular, we present a sequential fixing algorithm, an approximation algorithm, as well as a randomized algorithm, and derive several approximation ratios and performance bounds. 

\subsection{Sequential Fixing Algorithm}

To solve the above problem, we first define an indicator variable $x_{m,n}$ as 
\begin{eqnarray}\label{eq:xmn}
x_{m,n}=\left\{\begin{array}{l l}
$1$, & \mbox{if user $n$ is connected to BS $m$}\\
$0$, & \mbox{otherwise}.
\end{array}\right. 
\end{eqnarray}
Then we reformulate the  problem as follows:
\begin{eqnarray}\label{eq:MILP}
  \min && T  \\
  \mbox{s.t.} && \sum_m x_{m,n}=1, \; \mbox{for all users} \nonumber\\
              && \sum_n t_{m,n}x_{m,n}\le T, \; \mbox{for all BS's} \nonumber\\
              && x_{m,n} \in \{0,1\}, \; \mbox{for all $n\in\mathcal{A}_m$} \nonumber\\
              && x_{m,n}=0, \; \mbox{for all $n\notin\mathcal{A}_m$}. \nonumber
\end{eqnarray}
In the formulated problem~(\ref{eq:MILP}), all the indicator variable $x_{m,n}$'s are binary, while $T$ is a real variable. Thus it is a mixed integer linear programming problem~\cite{Kleinberg05}, denoted by MILP, which is usually NP-hard.


The original MILP is next relaxed to a linear programming (LP) problem, denoted as RLP. Specifically, we allow binary variable $x_{m,n}$'s to take real values in $[0,1]$. 
Then, the MILP problem can be converted into RLP as follows: 
\begin{eqnarray}\label{eq:LP}
  \min && T  \\
  \mbox{s.t.} && \sum_m x_{m,n}=1, \; \mbox{for all users} \nonumber\\
              && \sum_n t_{m,n}x_{m,n}\le T, \; \mbox{for all BS's} \nonumber\\
              && x_{m,n}\ge 0, \; \mbox{for all $n\in\mathcal{A}_m$} \nonumber\\
              && x_{m,n}=0, \; \mbox{for all $n\notin\mathcal{A}_m$}. \nonumber
\end{eqnarray}
Since the sum of $x_{m,n}$'s is already upper bounded by $1$ in the first constraint, we remove the upper bounds of $x_{m,n}$'s in the third constraint of MILP. Obviously, the solution to the RLP problem is a lower bound of the original MILP problem because it is obtained by expanding the solution space. Unfortunately, it is usually an infeasible solution to the original MILP problem. Therefore, we develop a sequential fixing (SF) algorithm~\cite{Hou08} to find a feasible solution to the MILP problem, which is presented in Algorithm~\ref{tab:SeqFixAlgo}. 


\begin{algorithm} [!t]
\small
\SetAlgoLined
   Initialize $\mathcal{N}=\{1,\cdots,N\}$ \;
   Relax $x_{m,n}$ to real numbers \;
   \While{$\mathcal{N}$ is not empty}{
   	Solve the RLP problem \;
   	Find $x_{m',n'}$ that is the closest to integer \;
   	$x_{m',n'}=\min_{n\in \mathcal{A}_m\cap\mathcal{N}}\{x_{m,n},1-x_{m,n}\}$ \;
   	Set $x_{m',n'}$ to the closest integer \;
   	\eIf{$x_{m',n'}$ is set to $1$}{
   		Set $x_{m,n'}=0$ for all $m \neq m'$ \;
   		Remove $n'$ from $\mathcal{N}$ \;
   	}{
   		Remove $n'$ from $A_{m'}$ \;
   	}   	
   }
\caption{Sequential Fixing for Cell Association}
\label{tab:SeqFixAlgo}
\end{algorithm}

Algorithm~\ref{tab:SeqFixAlgo}, we solve the RLP problem iteratively. During each iteration, we find the $x_{m',n'}$ that has the minimum value for ($x_{m,n}-0$) or ($1-x_{m,n}$) among all fractional $x_{m.n}$'s, and round it up or down to the nearest integer. Setting $x_{m',n'}$ to $1$ means user $n'$ is connected to BS $m'$. Therefore, user $n'$ cannot be connected to any other BS's and the rest of $x_{m,n'}$'s are set to $0$, for all $m$. This procedure repeats until all the $x_{m,n}$'s are fixed. 

The complexity of SF 
depends on the specific LP algorithm. With Karmarkar's algorithm, the worst-case polynomial bound for solving LP problems is $O({n_v}^{3.5}L_b)$, where $n_v$ is the number of variables and $L_b$ is the number of bits of input to the algorithm. We have the following proposition. 

\begin{propo}
The computational complexity of the sequential fixing algorithm is $O((MN)^{4.5}L_b)$.
\end{propo}
\begin{IEEEproof}
The number of binary variables in MILP is at most $MN$, so the number of loops in sequential fixing problem is at most $MN$. In each iteration, the complexities of Steps $4$, $5$ and the rest of the steps are $O((MN)^{3.5}L_b)$, $O(MN)$ and $O(1)$, respectively. Besides, in each iteration, the number of variables is reduced by $1$. Therefore, the complexity of SF
is given by
$\sum_{i=1}^{MN} O((MN-i+1)^{3.5}L_b) =\sum_{i=1}^{MN} O(i^{3.5}L_b)=O((MN)^{4.5}L_b)$.
Therefore, the complexity of SF
is upper bounded by $O((MN)^{4.5}L_b)$.
\end{IEEEproof}

\subsection{Approximation Algorithm} \label{subsec:aa}

Although the sequential fixing algorithm can solve the MILP problem within polynomial time,  its complexity may be high even for small femtocell networks. In this section, we propose an approximation algorithm with low complexity to solve the MILP problem. Before we introduce the approximation algorithm, we first give the lemma below.
\begin{lemma}\label{lemma:LB}
The optimal solution, denoted by $T^\ast$, to the MILP problem is lower bounded by $T^\ast\ge \frac{1}{M}\sum_{n=1}^N \underline{t}_n$ where $\underline{t}_n=\min_{m\in \mathcal{B}_n} t_{m,n}$.
\end{lemma}
\begin{IEEEproof}
Given the optimal allocation $\mathcal{C}^{\ast}_m$ for BS $m$, we have $T^\ast=\max_m \sum_{n\in \mathcal{C}^{\ast}_m}t_{m,n}$. Then we have
\begin{eqnarray}
T^\ast\ge\max_m \sum_{n\in \mathcal{C}^{\ast}_m}\underline{t}_n\ge\frac{1}{M}\sum_{m=1}^M \sum_{n\in \mathcal{C}^{\ast}_m}\underline{t}_n=\frac{1}{M}\sum_{n=1}^N\underline{t}_n. \nonumber 
\end{eqnarray}
The first inequality is due to the definition of $\underline{t}_n$. The second inequality is due to the fact that the maximum value is always greater than the mean value. The last equality is because all users have to be connected to one of the BS's and $\cup_{m=1}^M\mathcal{C}^{\ast}_m$ is the set of all users. 
\end{IEEEproof}

Intuitively, the maximum total service time is at least the service time of any one user. We have the following lemma.
\begin{lemma}\label{lemma:LB2}
The optimal solution, denoted by $T^\ast$, to the MILP problem is lower bounded by $T^\ast\ge \max \underline{t}_n$, where $\underline{t}_n=\min_{m\in \mathcal{B}_n} t_{m,n}$.
\end{lemma}

These lemmas will be used in analyzing the approximation ratio of the proposed approximation algorithms, which are presented in following subsections.

\smallskip
\subsubsection{Rounding Approximation Algorithm} \label{subsubsec:raa}

To ensure required SINR for each user, $\mathcal{B}_n$ should not include all the FBS's in a real femtocell network. For example, some faraway FBS should not be considered by a user. Thus, we can use a threshold $\rho$ to obtain the subsets $\mathcal{A}_m$ and $\mathcal{B}_n$ ($\mathcal{A}_m$ will be updated when $\mathcal{B}_n$ is determined).
\begin{eqnarray} \label{eq:rhodef}
\mathcal{B}'_n=\mathcal{B}_n \cap \left(\left\{m|t_{m,n}/\underline{t}_n \le \rho\right\}\right), \;\;
\mathcal{A}'_m=\{n|m\in \mathcal{B}'_n\}. 
\end{eqnarray}
Usually only a limited number of FBS's will be taken into consideration for a user. After we adopt this threshold, not only users' SINR requirements will be satisfied, but also the computational complexity will be greatly reduced.

Once $\mathcal{A}'_m$ and $\mathcal{B}'_n$ are determined, the following relaxed LP problem can be solved by any LP solver.
\begin{eqnarray}\label{eq:RLP}
 \min && T \\
 \mbox{s.t.} && \sum_m x_{m,n}=1, \; \mbox{for all users} \nonumber\\
      && \sum_n t_{m,n}x_{m,n}\le T, \; \mbox{for all BS's} \nonumber\\
      && x_{m,n}\ge 0, \; \mbox{for all $n\in\mathcal{A}'_m$} \nonumber\\
      && x_{m,n}=0, \; \mbox{for all $n\notin\mathcal{A}'_m$}. \nonumber
\end{eqnarray}
We denote the solution obtained by solving this RLP program by $T$. Since $x$-variables are allowed to take fractional values, we have $T \le T^\ast$. 

Without sequentially fixing these fractional values, we adopt a rounding method from~\cite{Shmoys93} to obtain a feasible solution for the MILP problem. In this rounding method, a bipartite graph is constructed according to the RLP solution, which is constructed as a undirected bipartite graph $G(\mathcal{A}\cup \mathcal{B},E)$. In the disjoint set $\mathcal{A}$, each node represents a user $n$, while the other disjoint set $\mathcal{B}$ consists of BS nodes. We create $k_m=\lceil \sum_n x_{m,n} \rceil$ nodes in $\mathcal{B}$ for BS $m$ and these node are denoted by $\{ b_{m,1},b_{m,2},\cdots,b_{m,k},\cdots,b_{m,k_m}\}$. The edges are determined in the following way. For BS $m$, we sort the users in the order of non-increasing service time $t_{m,n}$ and the users are renamed $\{u_1,u_2,\cdots\}$. Let $X_{m,u_j}=\sum_{i=1}^{j} x_{m,u_i}$. For each BS, we divide the users associated to it into $k_m$ groups, as $G_1, G_2, \cdots, G_{K_m}$. User $u_j$ will be included in group $k$ ($1 \le k \le k_m$) if $k-1 < X_{m,u_j} \le k$ or $k-1 \le X_{m,u_{j-1}} < k$. If a user $u_j$ is included in two groups, the association $x$-variables need to be adjusted, such that $x'_{b_{m,k},u_j}=X_{m,u_j}-k+1$ and $x'_{b_{m,k-1},u_j}=x_{m,u_j}-x'_{b_{m,k},u_j}$. Then we insert edges between BS node $b_{m,k}$ and all the user nodes in group $k$. Now the bipartite graph is created and we next find a maximum matching $\mathcal{M}$ from each user to nodes in the other disjoint set. This maximum matching $\mathcal{M}$ indicates a feasible solution for MILP problem: for each edge $(n,b_{m,k})$ in $\mathcal{M}$, we associate user $n$ to BS $m$.

Let $T_{(b_{m,k})}$ denote the total service time at node $b_{m,k}$ before the matching operation and $T'_{(b_{m,k})}$ the total service time at node $b_{m,k}$ obtained by the above rounding method. We have the following lemma.
\begin{lemma}\label{lemma:Tint}
For each node $b_{m,k}$, where $ k_m \ge k>1$, we have $T_{(b_m,k-1)} \ge T'_{(b_m,k)}$.
\end{lemma}
\begin{IEEEproof}
First, observe that the minimum service time in group $(k-1)$ will be always no less than the maximum service time in group $k$, because we sort the users according to their service time in the non-increasing order.

According the above bipartite graph construction, for any $k<k_m$, we have $\sum_{i\in G_k} x'_{b_{m,k},u_i}=1$; for $k=k_m$, we have $\sum_{i\in G_k} x'_{b_{m,k},u_i}\le1$.

$T'_{(b_m,k)}$ will be no greater than the maximum service time in group $k$ and will thus be no greater than the minimum service time in group $(k-1)$, which is less than $\sum_{i\in G_{k-1}} x'_{b_{m,k-1},u_i} t_{m,u_i}$. Since $T_{(b_m,k-1)}=\sum_{i\in G_{k-1}} x'_{b_{m,k-1},u_i} t_{m,u_i}$, consequently, we have the conclusion that $T_{(b_m,k-1)} \ge T'_{(b_m,k)}$.
\end{IEEEproof}

Now we show that the solution produced by this rounding approximation algorithm is at most $(\rho+1)$ times greater than the optimal solution. 
\begin{theorem}
The approximation algorithm based on linear programming and the rounding method ensures a 
$(\rho+1)$-approximation of the optimal solution.
\end{theorem}
\begin{IEEEproof}
For each BS $m$, we create $k_m$ nodes for it and there are $k_m$ corresponding groups of user nodes adjacent to them. Thus the total service time is $\sum_{k=1}^{k_m} T'_{(b_m,k)}$.

According to Lemma~\ref{lemma:Tint}, we have $T_{(b_m,k-1)} \ge T'_{(b_m,k)}$ for $k_m \ge k>1$. 
It follows that 
\begin{eqnarray}
\sum_{k=2}^{k_m} T'_{(b_m,k)} \le \sum_{k=1}^{k_m-1} T_{(b_m,k)} \le \sum_{k=1}^{k_m} T_{(b_m,k)} \le T. \nonumber
\end{eqnarray}
In the first group, the maximum load will be the maximum service time of users associated with $m$. 
According to  Lemma~\ref{lemma:LB2} and the definition of $\rho$ in~(\ref{eq:rhodef}), we have $T'_{(b_m,1)}  \le \max t_{m,n} \le \rho  \max \underline{t}_n \le \rho T^\ast $.
%
Then, the total service time on any BS computed by our association algorithm will be $\sum_{k=1}^{k_m} T'_{(b_m,k)} \le \rho T^\ast + T \le (\rho+1)  T^\ast$. The last inequality was due to $T \leq T^\ast$, since $T$ is the solution of the relaxed problem~(\ref{eq:RLP}). 
Our proof is complete.
\end{IEEEproof}

The complexity to compute a maximum matching is $O(VE)$, where $V$ and $E$ are the number of nodes and edges, respectively. Since we only need to run the matching algorithm once to obtain the association relationship, 
the total computational complexity of this algorithm is $O((MN)^{3.5}L_b)$, which is better than that of the sequential fixing algorithm.

\begin{propo}
The computational complexity of the rounding approximation algorithm is $O((MN)^{3.5}L_b)$.
\end{propo}

\smallskip
\subsubsection{Greedy Approximation Algorithm}

We next present a low complexity approximation algorithm, where the BS with the lowest load is greedily chosen and the user whose completion time at this BS is the smallest is assigned to this BS. 

By abuse of notation, we define $\rho_{m,n}=t_{m,n} / \underline{t}_{n}$ and $\rho=\max_{\{m,n\}}\rho_{m,n}$, which will be used in the optimality analysis. The greedy approximation algorithm is presented in Algorithm~\ref{tab:ApproxAlgo}. 
 In Step $4$, we find the candidate BS 
for users that has the minimum $T_m$. Then we pick the user who has the minimum $T_{m,n}$ at the chosen BS in Step $5$.   
Obviously, the computational complexity of the approximation algorithm is $O(MN)$, which is much lower than that of sequential fixing.
 

\begin{algorithm} [!t]
\small
\SetAlgoLined
   Initialize $T_m=0$ and $\mathcal{C}_m=\phi$ for all BS's \;
   Set the user set $\mathcal{N}=\{1,\cdots,N\}$ \;
   \While{$\mathcal{N}$ is not empty}{
   	Find the BS $m'$ that has the minimum $T_m$: 
   	$m'=\argmin_{m\in (\cup_{n\in\mathcal{N}}\mathcal{B}_n)}T_m$ \;
   	Find the user $n'$ that has the minimum $t_{m',n}$:
   	$n'=\argmin_{n\in\{\mathcal{A}_{m'}\cap\mathcal{N}\}} t_{m',n}$ \;
   	Set $\mathcal{C}_{m'}=\mathcal{C}_{m'}\cup\{n'\}$ \;
   	Set $T_{m'}=T_{m'}+t_{m',n'}$ \;
   	Set $\rho_{m',n'}=\frac{t_{m',n'}}{\underline{t}_{n'}}$ \;
   	Remove $n'$ from $\mathcal{N}$ \;
   }
\caption{Greedy Approximation Algorithm for Cell Association}
\label{tab:ApproxAlgo}
\end{algorithm}

\begin{propo}
The computational complexity of the greedy approximation algorithm is $O(MN)$.
\end{propo}

We have the following lemma for the performance of the greedy approximation algorithm. 
\begin{lemma}\label{lemma:UB}
The greedy approximation algorithm solution, denoted by $T$, is upper bounded by $\frac{\rho}{M}\sum_{n=1}^N \underline{t}_n + \rho T^\ast$. 
\end{lemma}

\begin{IEEEproof}
We first consider the open access strategy where each user can connect to any of the BS's. In the $l$-th iteration in Algorithm~\ref{tab:ApproxAlgo}, we choose the BS with the minimum $T_m$ in Step $4$. Thus we have
\begin{eqnarray}
T_{m'}^{(l-1)} \hspace{-0.075in}&\le&\hspace{-0.075in} \frac{1}{M}\sum_{m=1}^M T_m^{(l-1)}=\frac{1}{M}\sum_{m=1}^M\sum_{n\in \mathcal{C}_m^{(l-1)}}t_{m,n} \nonumber  \\
&\hspace{-0.075in}=&\hspace{-0.075in} \frac{1}{M}\sum_{m=1}^M\sum_{n\in \mathcal{C}_m^{(l-1)}}\rho_{m,n}\underline{t}_n \le \frac{\rho^{(l-1)}}{M}\sum_{m=1}^M\sum_{n\in \mathcal{C}_m^{(l-1)}}\underline{t}_n, \nonumber
\end{eqnarray}
where $\rho^{(l-1)}=\max_{\{m,n\in \mathcal{C}_m^{(l-1)}\}}\rho_{m,n}$. Note that $\mathcal{C}_m^{(l-1)}$ is set of users that have been assigned to BS $m$ in the ($l-1$)-th iteration. 

In Step $5$, we pick user $n'$ and let user $n'$ connect to BS $m'$. Since $\rho^{(l)}$ will always be greater than $\rho^{(l-1)}$ and according to Lemma~\ref{lemma:LB2}, we have
\begin{eqnarray}
 T_{m'}^{(l-1)}+t_{m',n'}\le\frac{\rho^{(l)}}{M}\sum_{m=1}^M\sum_{n\in \mathcal{C}_m^{(l)}}\underline{t}_n + \rho^{(l)} \underline{t}_n'. \nonumber
\end{eqnarray}
The algorithm stops after $N$ iterations. Since $T^{(l+1)}=\max\{T^{(l)},T_{m'}^{(l)}+t_{m',n'}\}$ and $T^{(0)}=0$, we conclude that
\begin{eqnarray}
T&=&T^{(N+1)}=\max\left\{T^{(N)},T_{m'}^{(N)}+t_{m',n'}\right\} \nonumber\\
&\le&\frac{\rho}{M}\sum_{m=1}^M\sum_{n\in \mathcal{C}_m}\underline{t}_n + \rho T^\ast
=\frac{\rho}{M}\sum_{n=1}^N \underline{t}_n+ \rho T^\ast. \nonumber
\end{eqnarray}
With the closed access stragegy, we set $t_{m,n}=\infty$, for BS $m$ that user $n$ cannot connect to, for all $m$, $n$. The proof follows the same procedure and we have the same conclusion. 
\end{IEEEproof}

Combining Lemmas~\ref{lemma:LB} and~\ref{lemma:UB}, we have the following theorem regarding the performance of Algorithm~\ref{tab:ApproxAlgo}.
\begin{theorem}\label{lemma:Bounds}
The greedy approximation algorithm in Algorithm~\ref{tab:ApproxAlgo} 
ensures a $(2\rho)$-approximation of optimal solution.
\end{theorem}
\begin{IEEEproof}
The proof is straightforward. We have
\begin{eqnarray}
 T^\ast\le T\le\frac{\rho}{M}\sum_{n=1}^N\underline{t}_n + \rho T^\ast \le 2 \rho T^\ast, \nonumber
\end{eqnarray}
where $T^\ast$ is the optimal solution and $T$ is the greedy approximation algorithm solution. Note that unlike in Section~\ref{subsubsec:raa}, we have $T^\ast\le T$ since there is no relaxation here. 
\end{IEEEproof}

From Theorem~\ref{lemma:Bounds}, 
$\rho$ is an important parameter to the performance of the greedy approximation algorithm. The smaller the $\rho$, the smaller the optimality gap. 
In order to make the greedy approximation algorithm solution more competitive, we only allow users to choose from a subset $\mathcal{B}_n$ of the original BS set. Then we have the new subsets $\mathcal{B}'_n$ and $\mathcal{A}'_m$ as 
\begin{equation}\label{eq:SetAB}
\mathcal{B}'_n=\mathcal{B}_n \cap \left( \left\{ m|\frac{t_{m,n}}{\underline{t}_n} \le \Gamma \right\}\cup \{1\} \right), 
\mathcal{A}'_m=\{n|m\in \mathcal{B}'_n\}, 
\end{equation}
where $\Gamma$ is a predefined threshold and $\{1\}$ is the index of the MBS. $\Gamma$ can also be used to indicate the SINR requirement of users. 
The set $\mathcal{A}_m$ is replaced by $\mathcal{A}'_m$ accordingly. This way, the greedy approximation algorithm solution will be 
\begin{equation}
  T^\ast \leq T \leq 2 \Gamma T^\ast.
\end{equation} 

\subsection{Randomized Algorithm}

Both the rounding and greedy approximation algorithms are centralized algorithms that require frequent CSI updates. 
In this section, we introduce a randomized algorithm for the cell association problem. With the randomized algorithm, each user $n$ randomly chooses a subset of $\mathcal{B}_n$ to connect to. Once the subsets are determined, no information exchange is required among the users. We assume user $n$ connects to BS $m$ with probability $p_{m,n}$ and the expected service time for user $n$ on each BS is identical (i.e., by tuning the $p_{m,n}$'s), i.e., 
\begin{eqnarray}
p_{m,n}t_{m,n}=H_n, \; \mbox{for all} \; m\in \mathcal{B}_n. \nonumber
\end{eqnarray}
Since a BS with a smaller $t_{m,n}$ should have higher preference, we set $p_{m,n}$ proportional to $1/t_{m,n}$. Since each user has to choose a BS to connect to, we have $\sum_{m\in \mathcal{B}_n}p_{m,n}=1$ for all $n$. It follows that
\begin{eqnarray}\label{eq:Hn}
H_n=\frac{1}{\sum_{m\in\mathcal{B}_n}1/t_{m,n}}, \mbox{for all} \; n. 
\end{eqnarray}

The expected load on BS $m$, denoted by $\overline{T}_m$, is
\begin{eqnarray}\label{eq:TmE}
\overline{T}_m=\mathbb{E}[T_m]=\sum_{n\in\mathcal{A}_m} t_{m,n} p_{m,n}= \sum_{n\in\mathcal{A}_m} H_n, \mbox{for all} \; m.
\end{eqnarray}
Since users are randomly connected to the BS's, our objective 
is to minimize the maximum value of the expected load $\overline{T}_{max}$.
\begin{eqnarray}\label{eq:minTmE}
\min \mbox{\;\;\;} \overline{T}_{max}=\min \{ \max_{m} \mbox{\;\;\;} \overline{T}_m \}. 
\end{eqnarray}
It can be seen from~(\ref{eq:TmE}) that minimizing $\overline{T}_m$ is equivalent to reducing the number of users in $\mathcal{A}_m$. 

The randomized algorithm consists of two phases. In Phase I, we use a threshold $\Lambda$ to obtain the subsets $\mathcal{A}_m$ and $\mathcal{B}_n$.
\begin{eqnarray}\label{eq:SetAB2}
\mathcal{B}'_n=\mathcal{B}_n \cap (\{m|t_{m,n}\le \Lambda\}\cup \{1\}), 
\mathcal{A}'_m=\{n|m\in \mathcal{B}'_n\}.
\end{eqnarray}
Note that the subsets $\mathcal{A}'_m$ and $\mathcal{B}'_n$ are different from those defined in (\ref{eq:SetAB}): $\Lambda$ is the upper bound of service time $t_{m,n}$, while $\Gamma$ is the upper bound on the service time ratios. Thus we have all $t_{m,n}\le \Lambda$ for all $n$ and $n\in\mathcal{A}'_m$. Then we derive the upper bounds for $H_n$, $\overline{T}_m$ and $\overline{T}_{max}$ as
\begin{eqnarray}\label{eq:HnTm}
\left\{ \begin{array}{l}
H_n = \frac{1}{\sum_{m\in\mathcal{B}'_n}1/t_{m,n}} \le \frac{1}{\sum_{m\in\mathcal{B}'_n}1/\Lambda} = \frac{\Lambda}{|\mathcal{B}'_n|} \\ 
\overline{T}_m = \sum_{n\in\mathcal{A}'_m} H_n\le \frac{|\mathcal{A}'_m|}{\min_n|\mathcal{B}'_n|}\Lambda \\
\overline{T}_{max} = \max_m \overline{T}_m\le \frac{\max_m|\mathcal{A}'_m|}{\min_n|\mathcal{B}'_n|}\Lambda. 
        \end{array} \right. 
\end{eqnarray}
where $|\mathcal{A}'_m|$ and $|\mathcal{B}'_n|$ are the cardinalities of subsets $\mathcal{A}'_m$ and $\mathcal{B}'_n$, respectively. 
  
In Phase II, we aim to further reduce the sizes of $\mathcal{A}'_m$ and $\mathcal{B}'_n$. From~(\ref{eq:Hn}), we find that $H_{n'}$ gets increased when BS $m'$ is removed from set $\mathcal{B}'_{n'}$ and user $n'$ is removed from set $\mathcal{A}'_{m'}$ simultaneously. The increase, denoted by $\Delta_{m',n'}$, is given by
\begin{eqnarray}\label{eq:Delta}
\Delta_{m',n'} 
=\frac{1}{\sum_{m\in\mathcal{B}'_n} \hspace{-0.025in} 1/t_{m,n} \hspace{-0.025in}-\hspace{-0.025in} 1/t_{m',n'}} \hspace{-0.025in}-\hspace{-0.025in} \frac{1}{\sum_{m\in\mathcal{B}'_n} \hspace{-0.025in} 1/t_{m,n}}  \nonumber \\ 
\hspace{-0.0in} =\frac{1/t_{m',n'}}{(\sum_{m\in\mathcal{B}'_n} \hspace{-0.025in} 1/t_{m,n} \hspace{-0.025in}-\hspace{-0.025in} 1/t_{m',n'})(\sum_{m\in\mathcal{B}'_n} \hspace{-0.025in} 1/t_{m,n})}. 
\end{eqnarray}
For those BS's in the set $\{m|m\in \mathcal{B}'_{n'}, m\neq m'\}$, their $\overline{T}_m$'s become larger when BS $m'$ is removed from set $\mathcal{B}'_{n'}$ and user $n'$ is removed from set $\mathcal{A}'_{m'}$.  
On the other hand, $\overline{T}_{m'}$ is reduced by $H_{m',n'}$ according to~(\ref{eq:TmE}). 

The randomized algorithm is presented in Algorithm~\ref{tab:RandomAlgo}. In Step $2$, we find the users that each has more than one BS on their BS list $\mathcal{B}'_n$. Then from Step $5$ to Step $18$, we find the BS $m'$ with the largest $\overline{T}_{m'}$ and compute the possible maximum load $\overline{T}^{max}_{m',n}$ on BS's for all users that might be connected to BS $m'$, assuming user $n$ is removed from $\mathcal{A}''_{m'}$. In Step $19$, we pick user $n'$ with the minimum $\overline{T}^{max}_{m',n}$ value. If the value is less than the original $\overline{T}_{m'}$, we remove the BS-user pair $\{m',n'\}$ from sets $\mathcal{A}''_{m'}$ and $\mathcal{B}''_{n'}$. Otherwise, the algorithm is terminated. 
When the algorithm is executed, sets $\mathcal{A}''_{m'}$ and $\mathcal{B}''_{n'}$ are subsets of $\mathcal{A}'_{m'}$ and $\mathcal{B}'_{n'}$, respectively. Since the complexity from Step $5$ to Step $18$ is $O(MN)$ in the worst case, the complexity of the entire randomized algorithm is $O(M\times N^2)$. 

\begin{propo}
The computational complexity of the randomized algorithm is $O(M\times N^2)$.
\end{propo}

Finally, we have the following theorem on the performance of the randomized algorithm. 


\begin{theorem}
The maximum expected service time achieved by the randomized algorithm is upper bounded by
\begin{eqnarray}
\overline{T}_{max}\le \frac{\max_m|\mathcal{A}''_m|}{\min_n|\mathcal{B}''_n|}\times \max_n\max_{m\in\mathcal{B}''_n}t_{m,n}.
\end{eqnarray}
\end{theorem}
\begin{IEEEproof}
The proof is similar to the derivation of (\ref{eq:HnTm}), but the new upper bound of service time, $\max_n\max_{m\in\mathcal{B}''_n}t_{m,n}$, is used, instead of the service time bound $\Lambda$. 
\end{IEEEproof}

\begin{algorithm} [!t]
\small
\SetAlgoLined
   Initialize $\mathcal{A}''_m=\mathcal{A}'_m$, $\mathcal{B}''_n=\mathcal{B}'_n$ \;
   Set the user set $\mathcal{N}=\{n||\mathcal{B}''_n|>1\}$ \;
   Compute $\overline{T}_m$ according to (\ref{eq:TmE}) \;
   \While{$\mathcal{N}$ is not empty}{
   	 Find the BS $m'$ with $m'=\argmax_{m} \overline{T}_m$ \;
   	 \For{user $n$ in ($\mathcal{A}''_{m'}\cap\mathcal{N}$)}{
   	 	 Compute $\Delta_{m',n}$ according to (\ref{eq:Delta}) \;
   	 	 \For{$m=1$ to $M$}{
   	 	 	 \uIf{$m=m'$}{
   	 	 	 	 Set $\overline{T}'_{m'}=\overline{T}_{m'}-H_n$ \;
   	 	 	 }\uElseIf{$m$ in $\{m|m\in \mathcal{B}''_{n}\}$}{
   	 	 	   Set $\overline{T}'_m=\overline{T}_m+\Delta_{m',n}$ \;
   	 	 	 }\Else{
   	 	 	 	 Set $\overline{T}'_m=\overline{T}_m$ \;
   	 	 	 }
   	 	 }
   	 	 Set $\overline{T}_{m',n}^{max}=\max_m \overline{T}'_m$ \;
   	 }
   	 Find user $n'$ with $n'=\argmin_n \overline{T}_{m',n}^{max}$ \;
   	 \eIf{$\overline{T}_{m'}\ge\overline{T}_{m',n'}^{max}$}{
   	 	 Remove $m'$ from $\mathcal{B}''_{n'}$ and $n'$ from $\mathcal{A}''_{m'}$ \;
   	 	 Update all $\overline{T}_m$'s \;
   	 	 \If{$|\mathcal{B}''_{n'}| =1$}{
   	 	 	 Remove $n'$ from $\mathcal{N}$
   	 	 }
   	 }{
   	 	 The algorithm is terminated \;
   	 }
   }
\caption{Randomized Algorithm for Cell Association}
\label{tab:RandomAlgo}
\end{algorithm}

\section{Service Scheduling}\label{sec:SevSchd}

Once the cell associate problem is solved as in Section~\ref{sec:ProbSol}, we then study how to schedule the transmissions of multiple users connecting to the same BS. Since we assume the bandwidth $B$ is fully utilized for transmitting a user's data packet (see~(\ref{eq:Cmn1})), the packets are transmitted consecutively. 
We need to determine the service order of the users that are associated with the same BS. 

Consider a tagged BS to which $K$ users are connected. The user service times are $\{t_1,t_2,\cdots,t_K\}$. 
If the service order follows the user index, the average waiting time is given by
\begin{eqnarray}
\overline{T}_{wait}=\frac{1}{K}\sum_{n=1}^K \sum_{i=1}^n t_i.
\end{eqnarray}     
We have the following theorem to minimize the average waiting time $\overline{T}_{wait}$. 

\begin{theorem}
Given $K$ users with service times $\{t_1, t_2, \cdots, t_K\}$, the average waiting time is minimized when the users are served in the increasing order of their service times.
\end{theorem}
\begin{IEEEproof}
First we sort the users according to their service times in the increasing order. The ordered service times are denoted by $\{t'_1,\cdots,t'_K\}$. Consider two ordered users $i$ and $j$, where $1\le i<j \le K$. We have $t'_i \le t'_j$. If the positions of $i$ and $j$ are swapped, it is obvious that the waiting times of users from $1$ to $i-1$ and the users from $j$ to $K$ are not affected and remain the same values. 
However, the awaiting time for each user from $i$ to $j-1$ is increased by $t'_j-t'_i$. Therefore, we conclude that the average waiting time is the least when the users are served in the increasing order of their service times.
\end{IEEEproof}

\section{Performance Evaluation}\label{sec:PerfEva}

In this section, we evaluate the performance of the proposed cell association and service scheduling algorithms using MATLAB simulations. The channel models from~\cite{Moon10} are adopted in our simulations. 
The channel gain (in $dB$) from the BS's to users can be expressed as $10\log(G_{m,n})=-PL_m(d_{m,n})-u_m$, 
where $d_{m,n}$ is the distance from BS $m$ to user $n$, and $u_m$ is the shadowing effect, which is normally distributed with a zero mean and variance $\delta_m$. The simulation parameters are presented in Table~\ref{tb:Parameter}.
In the figures, each point in the average of $10$ simulation runs; we included $95\%$ confidence intervals as error bars to make the simulation results credible.

\begin{table} [!t]
\begin{center}
\caption{Simulation Parameters}
\begin{tabular}{l|l}
\hline
{\em Paramter}            & {\em Value} \\
\hline
Number of BS's            & $6$   \\
Total network bandwidth   & $10 \mbox{ MHz}$ \\
Transmit power of the MBS & $43 \mbox{ dBm}$  \\
Transmit power of the FBS & $31.5 \mbox{ dBm}$   \\
Path loss model for MBS   & $28+35\log_{10}(d)$\\
Path loss model for FBS   & $38.5+20\log_{10}(d)$ \\
Shadowing effect          & $6 \mbox{ dB}$\\
Packet length             & $1 \mbox{ KBytes}$ \\
Threshold $\rho$          & 5 \\
\hline
\end{tabular}
\end{center}
\label{tb:Parameter}
\vspace{-0.15in}
\end{table}

\begin{figure*}[!t]
     \begin{center}
     		\subfigure[Total service time vs. number of users]{%
           \label{fig:TotTimeOpen}
           \includegraphics[width=2.25in]{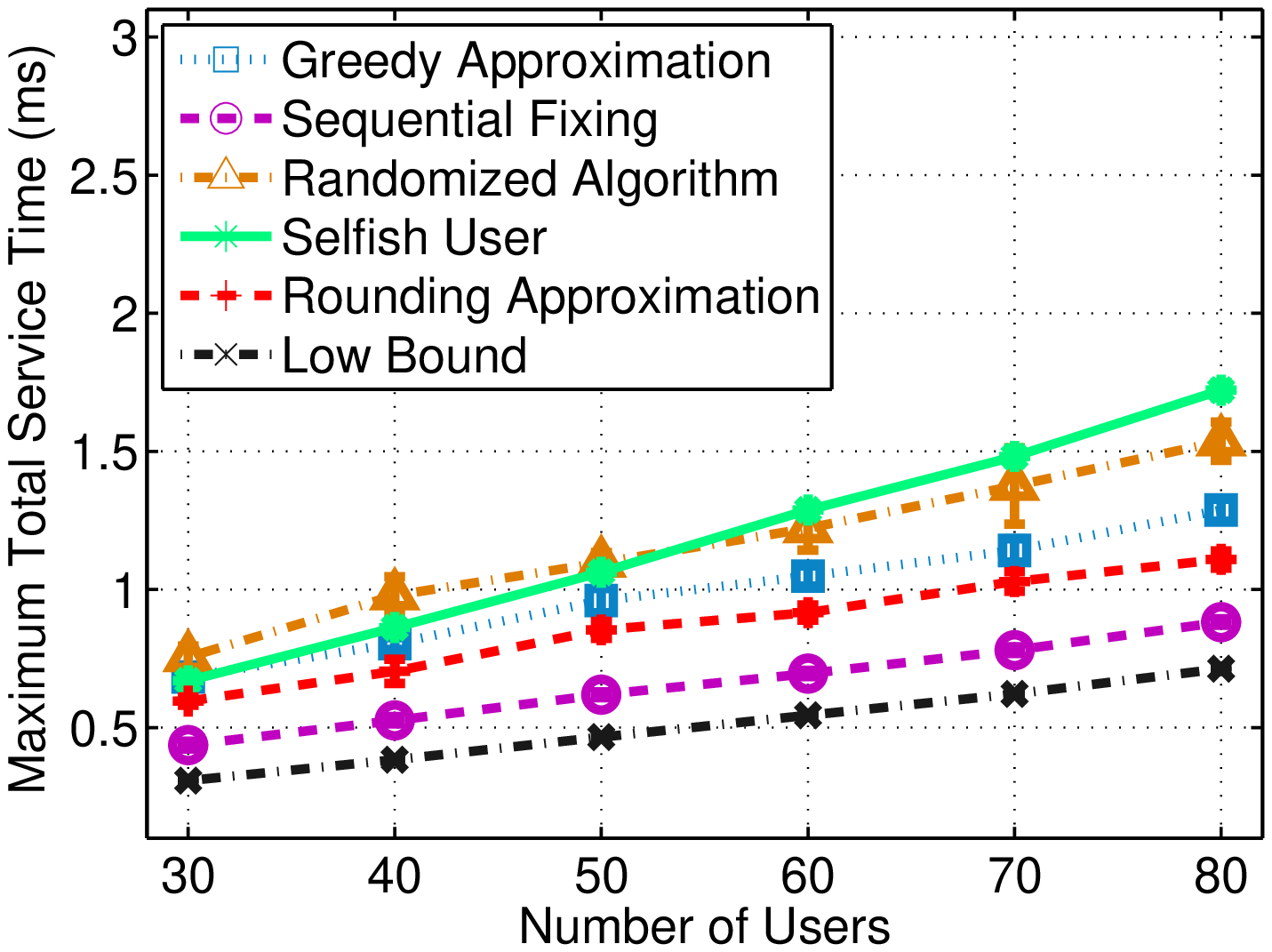}
        }
        \hfil
        \subfigure[Average waiting time vs. number of users]{%
           \label{fig:WaitTimeOpen}
           \includegraphics[width=2.25in]{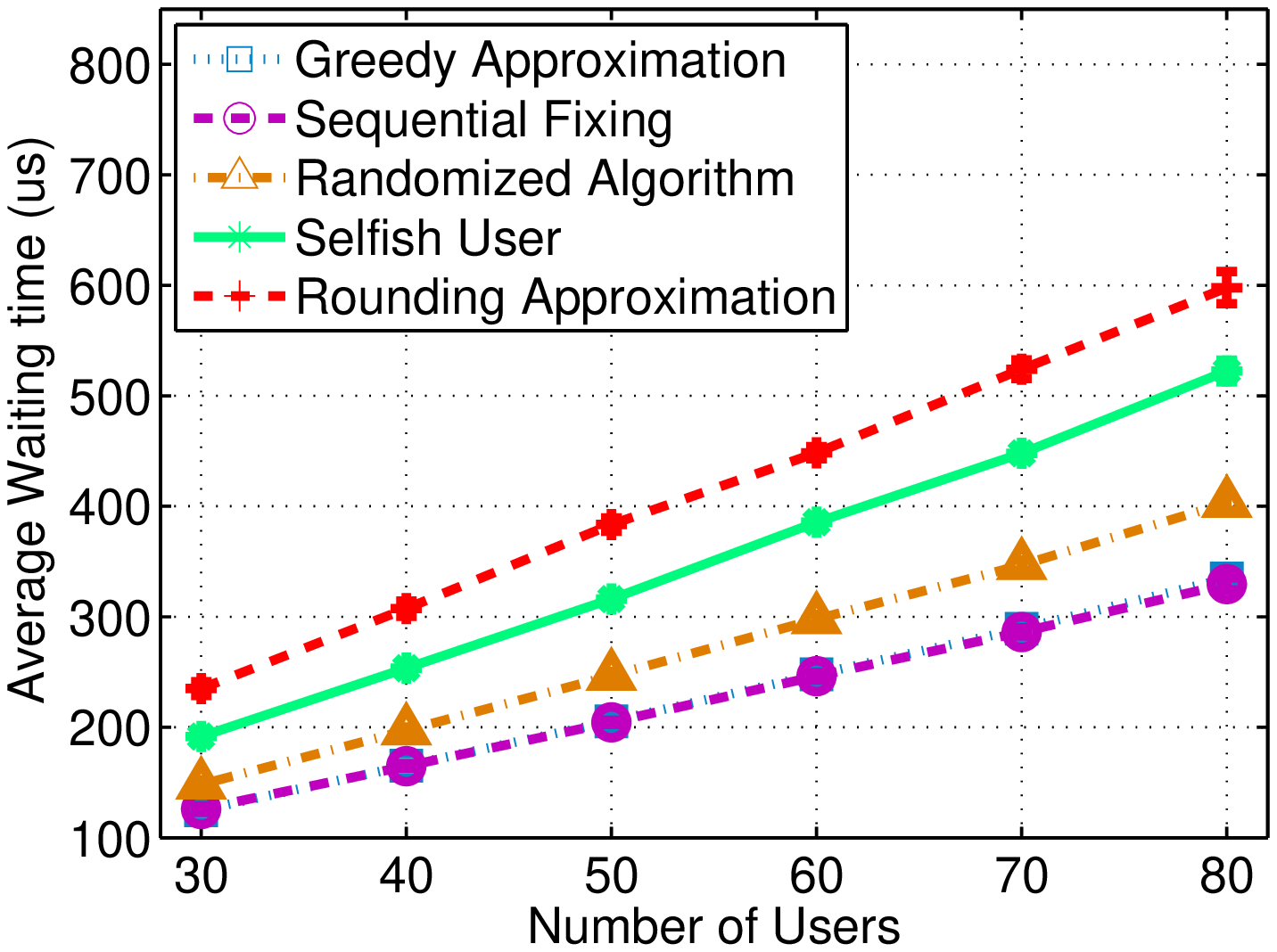}
        }
        \subfigure[Fairness vs. number of users]{%
           \label{fig:FairnessOpen}
           \includegraphics[width=2.25in]{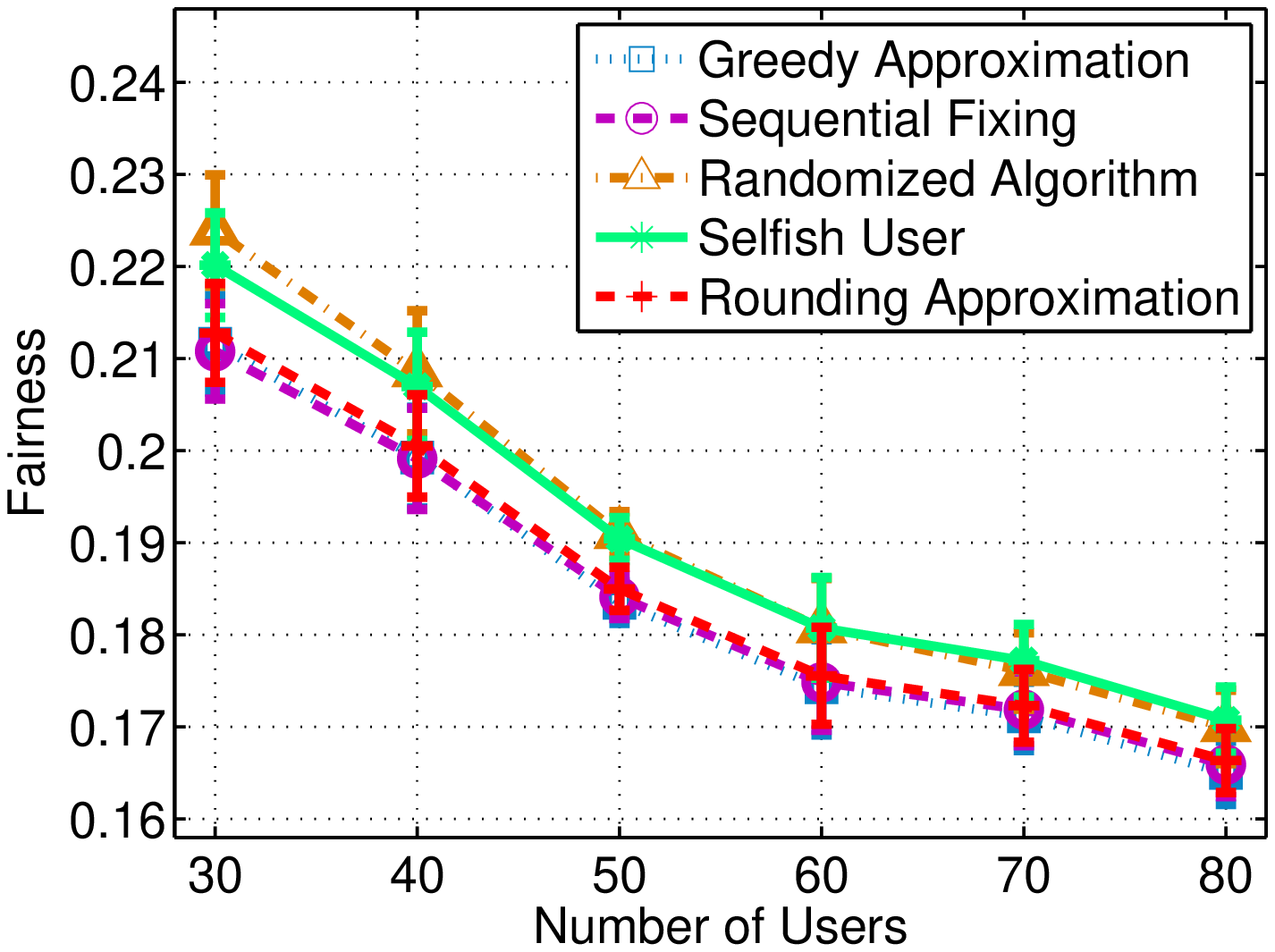}
        }          
    \end{center}
    \caption{Performance evaluation of the open access strategy.}%
\end{figure*}

We present simulation results for the following two scenarios: (i) open access femtocells; (ii) closed access femtocells. 
For comparison purpose, we also developed and simulated a selfish scheme and compared it with the proposed schemes. With the selfish scheme, every user simply chooses the BS with the best channel condition to connect to. 

\subsection{Open Access Strategy}

In the first scenario, there are $M=6$ BS's, i.e., one MBS and five FBS's. The number of users ranges from $30$ to $80$ with step size $10$. They are randomly located in network area. 
Each user can connect to one of the BS's.

\begin{table} [!t]
\begin{center}
\caption{Execution Times of the Proposed Algorithms under the Open Access Strategy (s)}
\setlength{\tabcolsep}{5pt}
\begin{tabular}{r|c|c|c|c|c|c}
\hline
No. users & 30&40&50&60&70&80 \\
\hline
\hline
Greedy &0.024&0.034&0.024&0.030&0.026&0.038\\
Approx. & & & & & & \\
\hline
Sequential&16.532&24.020&30.809&48.713&47.842&50.654\\
Fixing & & & & & & \\
\hline
Randomized &0.030&0.048&0.077&0.136&0.132&0.151\\
Algorithm & & & & & & \\
\hline
Selfish User&0.035&0.035&0.035&0.035&0.036&0.026\\
Scheme & & & & & & \\
\hline
Rounding &0.133&0.148&0.160&0.168&0.176&0.213\\
Approx. & & & & & & \\
\hline
\end{tabular}
\end{center}
\label{tb:Runtime_open}
\vspace{-0.15in}
\end{table}

\begin{figure*}[!t]
     \begin{center}
     		\subfigure[Total service time vs. number of users]{%
           \label{fig:TotTimeClose}
           \includegraphics[width=2.25in]{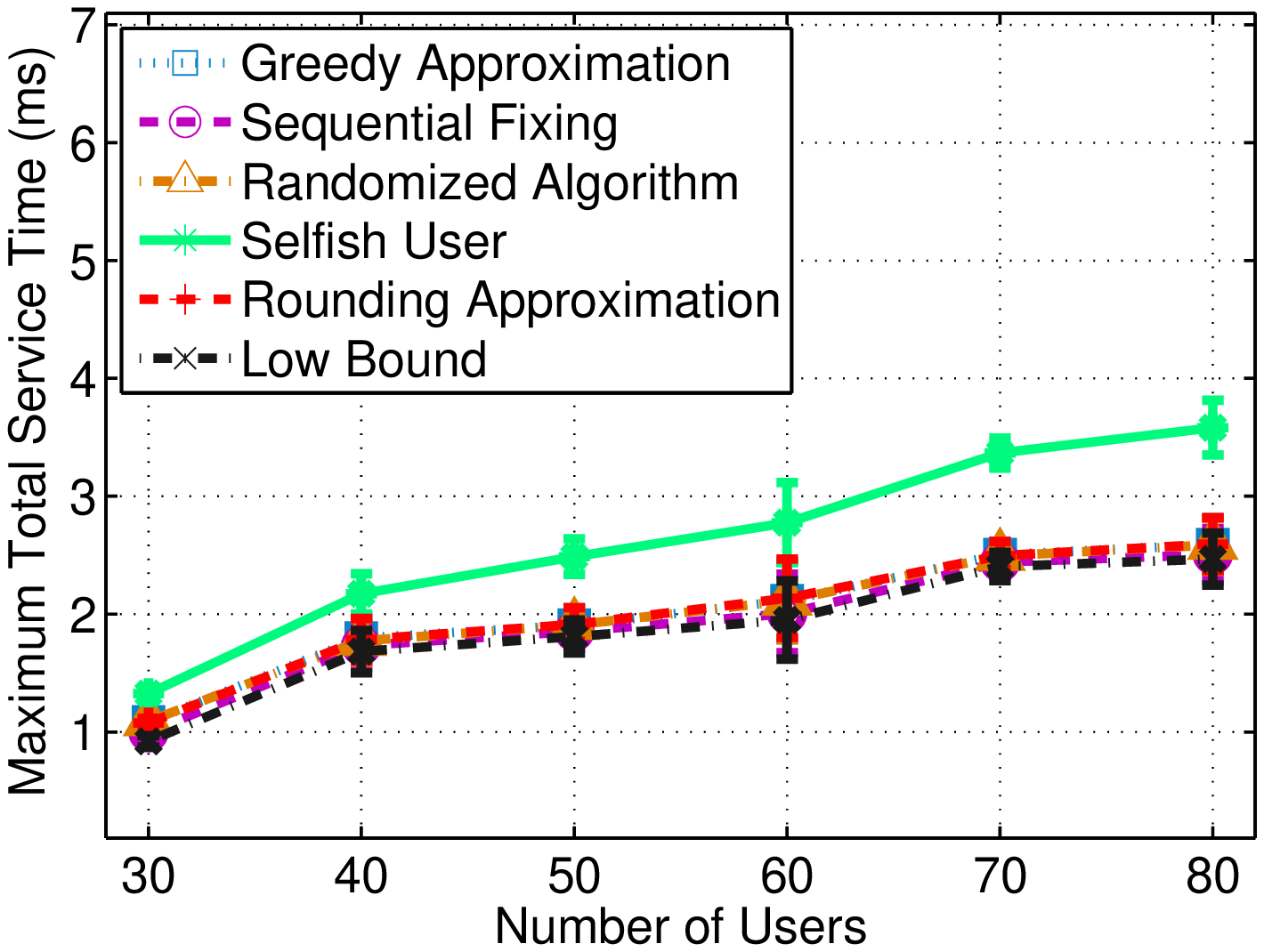}
        }
        \hfil
        \subfigure[Average waiting time vs. number of users]{%
           \label{fig:WaitTimeClose}
           \includegraphics[width=2.25in]{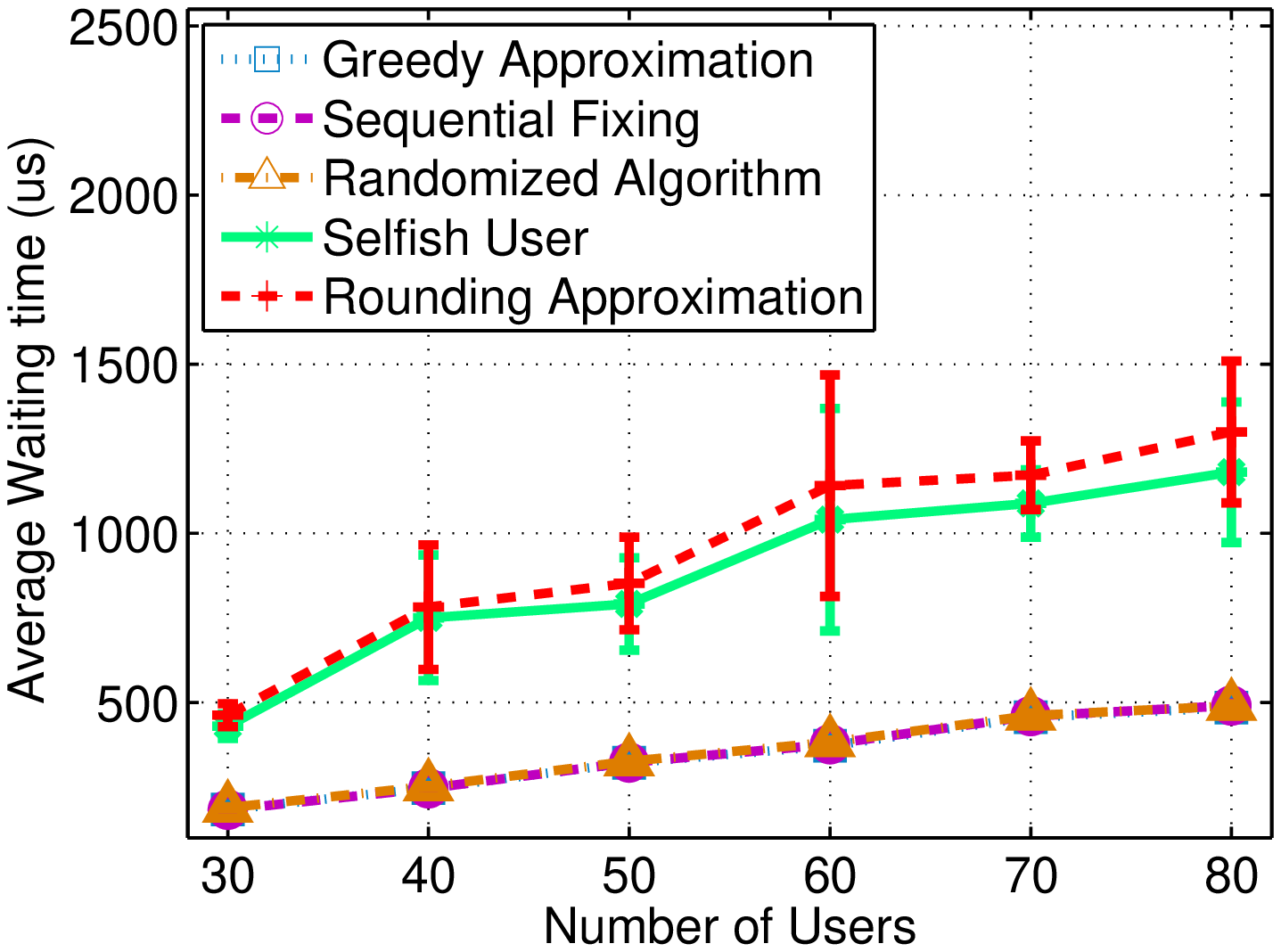}
        }   
        \subfigure[Fariness vs. number of users]{%
           \label{fig:FairnessClose}
           \includegraphics[width=2.25in]{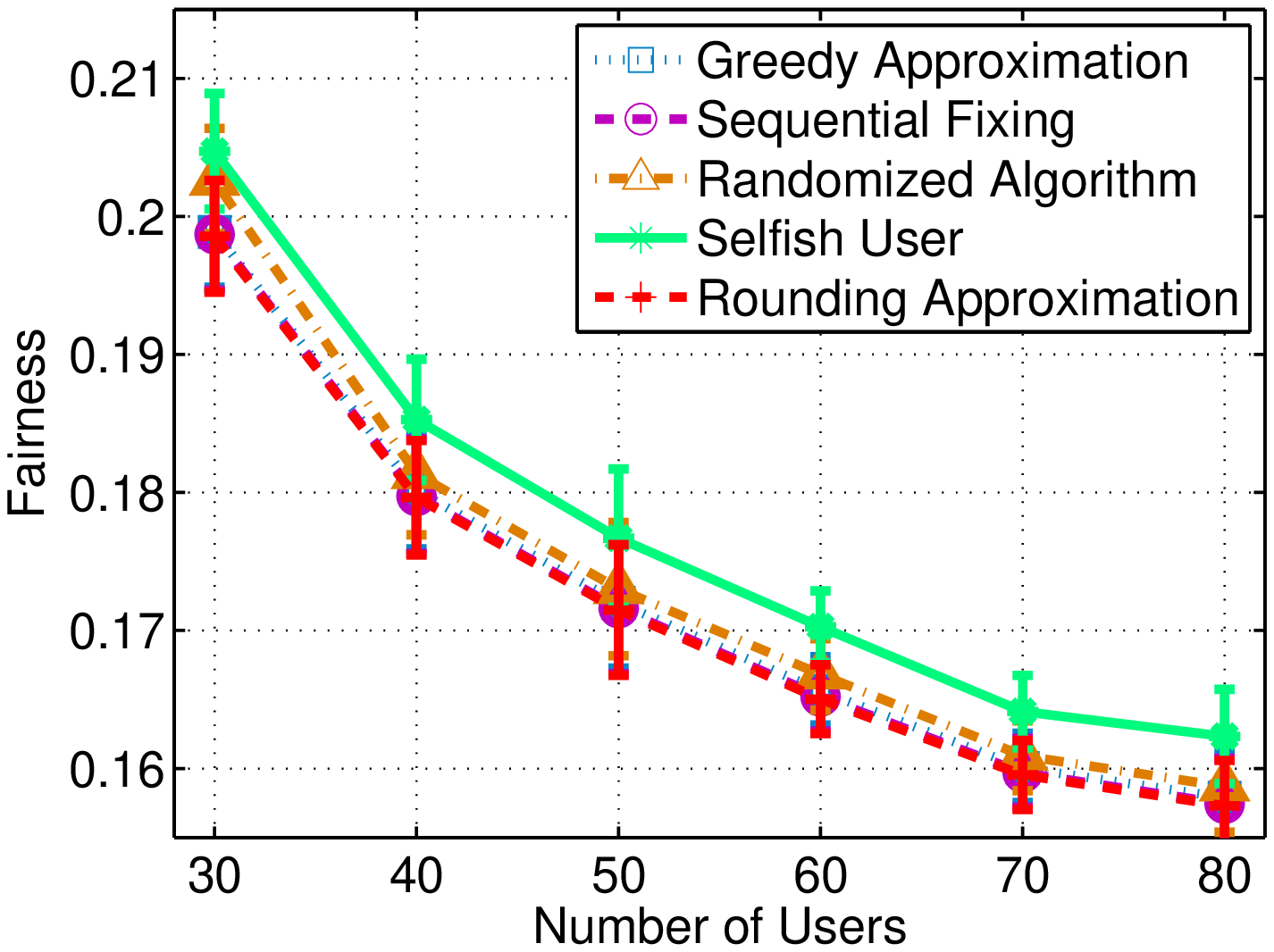}
        }  
    \end{center}
    \caption{Performance evaluation of the closed access strategy.}%
\end{figure*}

We first examine the impact of the number of users on total service time. In Fig.~\ref{fig:TotTimeOpen}, we plot the maximum total service time for the five algorithms along with the lower bound found by solving the relaxed LP. 
As expected, the more users, the more total service time on BS's. Except for the low bound, 
the sequential fixing algorithm achieves the smallest total service time. 
The rounding approximation algorithm has a slightly better performance than the greedy approximation algorithm 
and the result justifies the approximation ratio proven in Section~\ref{subsec:aa}. Both approximation algorithms always achieve lower load than both the randomized algorithm and the selfish scheme. We also observe that beyond $50$ users, all the proposed algorithms have lower service times than the simple selfish scheme. When number of users becomes larger, the simple selfish scheme becomes less competitive and the rounding approximation algorithm achieves almost $50\%$ less total service time in the case of $80$ users.

After cell association, users should be properly scheduled to get service in BS's to minimize average waiting time. In Fig.~\ref{fig:WaitTimeOpen}, we investigate the impact of the number of users on average waiting time. 
In the scheme of greedy approximation, randomized algorithm and sequential fixing, we use the service scheduling policy in Section~\ref{sec:SevSchd} to schedule users in BS's and obtain the corresponding waiting time. For comparison, we randomly schedule users in BS's in the selfish scheme and rounding approximation scheme. Intuitively, the larger the number of users, the larger the average waiting time. We can see from the figure that, the average waiting time obtained by the greedy approximation algorithm is very close to that by the sequential fixing algorithm, while without appropriate scheduling, the rounding approximation algorithm achieves the largest waiting time, which is almost twice as large as the waiting time achieved by greedy approximation algorithm.

To evaluate the fairness performance, we adopt Raj Jain's fairness index given by 
$\mathcal{J}(C_1,C_2,\cdots,C_N)=\frac{(\sum_{n=1}^NC_n)^2}{N\times\sum_{n=1}^NC_n^2}$, 
where $C_n$ is the network throughput for user $n$~\cite{Zhou13}. The value of the index ranges from $1/N$ (worst case) to $1$ (best case). It can be seen from Fig.~\ref{fig:FairnessOpen} that fairness indexes decrease when the number of users is increased. We notice that, the selfish scheme and the randomized algorithm achieve better fairness than the other three schemes. Figs.~\ref{fig:TotTimeOpen} and~\ref{fig:FairnessOpen} show that from operator's viewpoint, the selfish and the randomized schemes are not preferred since they produce less balanced load on BS's. From users's viewpoint, these two schemes may be appealing due to their fairness performance.

We list the execution times of the five schemes in Table~\ref{tb:Runtime_open}. We find the execution time increases as the number of users is increased. The selfish scheme always has the smallest execution time, while sequential fixing has the largest execution time. Although the rounding approximation algorithm can achieve smaller load on the BS's, its execution time is greater than that of the greedy approximation algorithm. 
This result also justifies the complexity analysis for the proposed schemes. The running time of the greedy approximation algorithm and the selfish scheme is always much smaller than other schemes and does not increase obviously with the number of users.
For the closed access simulations shown in Section~\ref{subsec:cas}, the execution times of the proposed algorithms are all much smaller than that shown in Table~\ref{tb:Runtime_open}, since the user list include fewer users in the closed access case. We omit these results for brevity.

\subsection{Closed Access Strategy}\label{subsec:cas}

We next investigate the second scenario with closed access femtocells. Now each FBS maintains a user list and only serves the listed users. Note that the MBS will always serve all the users inside its coverage. 


In Fig.~\ref{fig:TotTimeClose}, we evaluate the impact of the number of users on total service time. Intuitively, the total service time increases as the number of users. However, we find that it also depends on the user list at each FBS. In the simulation, we randomly choose the user set  $\mathcal{A}_m$ for BS $m$. Moreover, the user list at each FBS is further reduced due to the SINR threshold. Consequently, all the proposed algorithms achieve close performance in the closed access scenario. The total service time of the proposed algorithms is close to the low bound in closed access scenario. However, the performance of all the proposed algorithms is better than that of the selfish scheme, as we can see in Fig.~\ref{fig:TotTimeClose}. 

We next show the impact of the number of users on average waiting time in Fig.~\ref{fig:WaitTimeClose}. The scheduling policy setting is the same as that in the open access scenario. The result thus is also similar to the open access case that, the selfish scheme and the rounding approximation scheme achieve the largest waiting time.
Actually with proposed optimal service scheduling, the approximation algorithms will achieve as less waiting time as that of the sequential fixing scheme.

Finally, we plot the fairness indices in Fig~\ref{fig:FairnessClose}. The randomized algorithm, although not better than the selfish scheme, achieves the best performance in fairness than the other proposed schemes. Despite of its good performance in minimizing the maximum service time, the rounding approximation algorithm, is not competitive with respect to fairness. Due to the randomness of user lists at BS's, the confidential intervals are larger than those in the open access scenario.

\section{Conclusion}\label{sec:Conc}

In this paper, we investigated the problem of cell association and service scheduling in two-tier femtocell networks. 
We developed several algorithms and analyzed their performance. The sequential fixing algorithm achieves the best performance in total service time but it has a relatively high complexity. 
Then we presented two approximation algorithms with lower complexity and proven approximation ratios. We also proposed a randomized algorithm with a proven performance bound that requires the least information exchange among users. 
In addition, we addressed the service scheduling problem with an optimal solution. The proposed algorithms were validated with simulations in both open and closed access scenarios. 

%

\bibliographystyle{IEEEtran}
\bibliography{cell_assign_femto}

\end{document}